\documentclass[twocolumn,eqsecnum,aps]{revtex4}
\usepackage[dvips]{epsfig,color}
\usepackage{graphicx}
\usepackage{amsmath,amsfonts,amsbsy,amssymb}
\usepackage{tabularx}

\begin{document}

\title{Hyperbolic Supersymmetric Quantum Hall Effect}
\author{Kazuki Hasebe}
\affiliation{Department of General Education, Takuma National College of Technology, Takuma-cho, Mitoyo-city, Kagawa 769-1192, Japan \\
Email: hasebe@dg.takuma-ct.ac.jp}

\begin{abstract}

Developing a non-compact version of the supersymmetric Hopf map,  
we formulate the quantum Hall effect on a super-hyperboloid.  
Based on $OSp(1|2)$ group theoretical methods, we first analyze the one-particle Landau problem, and successively explore the many-body problem where Laughlin wavefunction, hard-core pseudo-potential Hamiltonian and topological excitations are derived. It is also shown that the fuzzy super-hyperboloid emerges in the lowest Landau level. 

\end{abstract}

\maketitle

\section{Introduction}

In the past several years, the understanding of higher dimensional formulations of the quantum Hall effect (QHE) has greatly progressed. 
The initial study of this direction may date back to the pioneer work of Haldane who 
formulated QHE on two-spheres more than two decades ago \cite{PRL511983}. 
Beyond the importance to the study of QHE itself, in a modern perspective, Haldane's QHE could be appreciated as a physical realization of fuzzy geometry on a curved manifold. However, reasonable higher dimensional generalizations of Haldane's model had not been found until the breakthrough of Zhang and Hu's four-dimensional QHE \cite{cond-mat/0110572}.
Since their discovery, many analyses have been devoted to further generalizations of QHE on other higher dimensional curved manifolds. Among them, QHEs on complex projective manifolds \cite{hep-th/0203264} and higher dimensional spheres \cite{cond-mat/0306045,hep-th/0310274} have been well explored accompanied with the developments of fuzzy geometry and matrix models \cite{hep-th/0606161}.

Since the previous investigations are mainly concerned with  
compact bosonic manifolds, there might be two successive directions to be pursued. 
One direction would be the exploration on non-compact manifolds.
With respect to hyperboloids, several works have already been reported, for the Landau problem \cite{AP173(1987)185,amp462005,NPB413(1994)735,hep-th/0602231v1} and for the QHE \cite{arXiv:hep-th/0505095,arXiv:hep-th/0605289,arXiv:hep-th/0605290,arXiv:0710.2292} as well. 
The other direction is the exploration on supermanifolds. 
Ivanov et al. launched the construction of the Landau model on compact supermanifolds, such as 
 supersymmetric complex projective spaces 
 \cite{hep-th/0311159}, super-flag manifolds 
\cite{hep-th/0404108}.
Independently, Hasebe and Kimura investigated Landau problem on a  supersphere   
\cite{hep-th/0409230} \footnote{There are two different definitions of supersphere, one of which is the coset $SU(2|1)/U(1|1)$ as used in \cite{hep-th/0311159} while the other is $UOSp(1|2)/U(1)$ in \cite{hep-th/0409230}. 
 In this paper, we adopt the latter definition to discuss the non-compact version of it.}.
Recently, particular properties of the supersymmetric (SUSY) Landau models are starting to be unveiled, such as non-anticommutative geometry in the lowest Landau level (LLL) \cite{hep-th/0311159,hep-th/0404108,hep-th/0409230,hep-th/0503162,hep-th/0510019}, 
enhanced SUSY in higher Landau levels 
\cite{hep-th/0503162,hep-th/0510019,hep-th/0612300,arXiv:0705.2249,arXiv:0806.4716}, and the  existence of negative norm states \cite{hep-th/0503162,hep-th/0510019}.
The remedy for the negative norm problem was implicitly suggested in Ref.\cite{hep-th/0503162}, and well developed in Refs.\cite{hep-th/0612300,arXiv:0705.2249,arXiv:0806.4716} by introducing the  appropriate metric in Hilbert space. 
Many-body problems on supermanifolds, which we call the SUSY QHE, 
have been also explored in Refs.\cite{hep-th/0411137,hep-th/0503162,arXiv:0705.4527,hep-th/0606007,arXiv:0710.0216}. 
The SUSY QHE was first formulated on a supersphere \cite{hep-th/0411137}, and next on a superplane \cite{hep-th/0503162,arXiv:0705.4527}.
Their corresponding bosonic ``body'' manifolds are, respectively, two-sphere and Euclidean plane, and both of them are maximally symmetric spaces with Euclidean signatures; the former has positive constant curvature, while the latter does zero constant curvature. 
Recently, it was also found that the set-up of the SUSY QHE was applicable to hole-doped antiferromagnetic quantum spin models \cite{SUSYAKLTpaper}.

In this paper, we explore a formulation of the QHE on a super-hyperboloid whose 
 body is the hyperboloid, which has negative constant curvature and is the last two-dimensional maximally symmetric space with a Euclidean signature. 
For the construction, we introduce a non-compact version of the SUSY Hopf map. The author believes this to be the first case where the non-compact SUSY Hopf map and its related materials are developed.
The hyperbolic formulation of the SUSY QHE would be interesting, also from fuzzy geometry and AdS/CFT points of view.  
 The hyperbolic SUSY QHE provides a nice physical realization of the fuzzy super-hyperboloid, 
and, interestingly, the fuzzy hyperboloid or fuzzy (Euclidean) $AdS^2$ naturally appears in the context of  AdS/CFT correspondence 
\cite{hep-th/0004072,hep-th/0005268}.
The hyperboloid SUSY QHE itself is closely related to the concept of holography. 
While on  spheres a natural definition of boundary does not exist, there is one on  hyperboloids or AdS spaces. Further, edge states in the QHE are described by the chiral CFT formalism \cite{WenPRL1990,StonePRB1990}, which reflect bulk properties governed by the Chern-Simons field theory. 
The bulk-edge correspondence in hyperbolic (SUSY) QHE is expected to demonstrate the concept of ``AdS/CFT'' in condensed matter physics.
  
In the first half of this paper, we formulate the QHE on a (bosonic) hyperboloid based on the non-compact Hopf map, and rederive several results reported in Refs.\cite{AP173(1987)185,amp462005,NPB413(1994)735,arXiv:hep-th/0505095,arXiv:hep-th/0605289,arXiv:hep-th/0605290}.
We provide new ingredients also, such as the pseudo-potential Hamiltonian and topological excitations. In the latter half, we extend the discussions to the super-hyperboloid case, where we explore the non-compact SUSY Hopf map, and construct a formulation of the hyperbolic SUSY QHE. 
The detailed organization of this paper is as follows.  In Sec.\ref{sectsu11group}, we briefly review basic properties of the $SU(1,1)$ group.
In Sec.\ref{sectsu11hopf}, the non-compact Hopf map is introduced.
The one-particle problem on the hyperboloid is discussed in Sec.\ref{oneparticleprob}.
The noncommutative geometry in the LLL is derived, and Hall relation is confirmed in Sec.\ref{sectnoncommutative}.
In Sec.\ref{sectqhe}, we discuss the many-body problem on the hyperboloid.  
From Sec.\ref{sectiosp11group} to Sec.\ref{sectsusyqhe}, with the use of the $OSp(1|2)$ super Lie group, we supersymmetrize the previous discussions. 
Sec.\ref{summarysection} is devoted to summary and discussions.
Several definitions related to supermatrix are given in Appendix \ref{defsinsuper}.
In Appendix \ref{sectlagrange}, the Lagrange formalism on the super-hyperboloid is provided. The irreducible representations of the $SU(1,1)$ group are summarized in Appendix \ref{appendirredrepsu11}.

\section{Preliminaries I}\label{sectsu11group}

\subsection{The $SU(1,1)$ Group and Algebra}

$SU(1,1)$ is topologically equivalent to a not-simply connected non-compact manifold $D\times S^1$ ($D$ represents a disk), and is isomorphic to several groups, 
\begin{equation}
SU(1,1)\simeq SL(2,R)\simeq Sp(2,R)
\end{equation}
and 
\begin{equation}
SU(1,1)/Z_2\simeq SO(2,1).
\end{equation}
The $SU(1,1)$ group element $g$ is defined so as to satisfy the relation
\begin{equation}
g^{\dagger}\sigma^3 g=\sigma^3,
\label{su11trans}
\end{equation}
with the constraint 
\begin{equation}
\text{det}(g)=1. 
\label{determinatg}
\end{equation}
When $g$ is expressed as  
\begin{equation}
g=
\begin{pmatrix}
u & v^* \\
v & u^*
\end{pmatrix},
\label{explicitgmatrix}
\end{equation}
the constraint (\ref{determinatg}) becomes  
\begin{equation}
u u^*-v v^*=1.
\end{equation}
 The inverse of $g$ is given by  
\begin{align}
&g^{-1}=\sigma^3 g^{\dagger}\sigma^3=
\begin{pmatrix}
u^* & -v^* \\
-v & u 
\end{pmatrix}
\nonumber\\
&~~~~~\neq g^{\dagger}=
\begin{pmatrix}
u^* & v^* \\
v & u 
\end{pmatrix}.
\end{align}
Since $SU(1,1)$ is a non-compact group, its unitary representation  is infinite-dimensional. (The irreducible representations of $SU(1,1)$ are summarized in 
Appendix \ref{appendirredrepsu11}, and detailed discussions can be found in Ref.\cite{am48(1947)}.) In this paper, we deal with non-unitary representation of the principal discrete series, and hence the generators are generally represented by non-Hermitian and finite dimensional matrices. 
The $SU(1,1)$ generators are given by 
\begin{equation}
s^a=\frac{1}{2}\kappa^a,
\end{equation}
where $\kappa^a$ are  
\begin{equation}
\kappa^1=i\sigma^1,~~\kappa^2=i\sigma^2,~~\kappa^3=\sigma^3.
\label{nonunitaryrepSU11}
\end{equation}
Here, $\sigma^a$ denote Pauli matrices; non-Hermitian matrices $\kappa^1$ and $\kappa^2$ are boost generators to $x$ and $y$ directions, respectively, while
 the Hermitian matrix $\kappa^3$ is the rotation generator on the $x\!-\!y$ plane.    
$s^a$ satisfy the algebra,
\begin{equation}
[s^a,s^b]=i\epsilon^{ab}_{~~~c} s^c, 
\label{su11algebraabouts}
\end{equation}
where $\epsilon^{abc}$ represents the three-rank antisymmetric tensor with $\epsilon^{123}=1$, and the indices are raised or lowered by the metric $\eta_{ab}=\eta^{ab}=(+,+,-)$. 
$-s_a$ also satisfy the $SU(1,1)$ algebra, and are related to $s^a$ as 
\begin{equation}
\sigma^3 s^a \sigma^3=-s_a.
\end{equation}
The Casimir operator is given by 
\begin{equation}
C=\eta_{ab}s^a s^b=s^1s^1+s^2s^2-s^3s^3,
\end{equation}
and its eigenvalues are 
\begin{equation}
C=-j(j-1)
\end{equation}
with $j=1,{3}/2,2,{5}/{2},\cdots$. It should be noticed that the Casimir index $j$ begins from $1$ not $0$. 
We summarize the properties of $\kappa^a$ for later convenience.
Their anticommutation relations are given by 
\begin{equation}
\{\kappa^a,\kappa^b\}=-2\eta^{ab},
\end{equation}
and then, with (\ref{su11algebraabouts}), 
\begin{equation}
\kappa^a\kappa^b=-\eta^{ab}+i\epsilon^{ab}_{~~c}\kappa^c.
\end{equation}
Their normalizations are 
\begin{equation}
tr(\kappa^a \kappa^b)=-2\eta^{ab}.
\label{sasbtrace}
\end{equation}
The completeness relation is  
\begin{equation}
4\eta_{ab}(\kappa^a)_{\alpha}^{~~\beta}(\kappa^b)_{\gamma}^{~~\delta}
=-2\delta_{\alpha}^{~~\delta}\delta_{\beta}^{~~\gamma}+\delta_{\alpha}^{~~\beta}\delta_{\gamma}^{~~\delta}.
\end{equation}

\subsection{Complex Representation}

The complex representation is given by 
\begin{equation}
\tilde{\kappa}^a\equiv -{\kappa^a}^*=\kappa_a^t,
\label{complexrepsu11}
\end{equation}
and related to the original representation by the unitary transformation 
\begin{equation}
\tilde{\kappa}^a=R^{\dagger}\kappa^a R,
\end{equation}
where $R=\sigma^1$. 
Then, with an $SU(1,1)$ spinor $\phi$, 
its charge conjugation is constructed as 
\begin{equation}
\phi_c=R^{\dagger}\phi^{*},
\end{equation}
and the Majorana condition $\phi_c=\phi$ is given by    
\begin{equation}
\phi=\sigma^1 \phi^*,
\end{equation}
or 
\begin{equation}
{\phi^1}^*={\phi^2},~~{\phi^2}^*=\phi^1.
\end{equation}
Without introducing the complex conjugation, the $SU(1,1)$ singlet is constructed as 
\begin{equation}
(R^{\dagger}\varphi^*)^{\dagger}\sigma^3\phi=\varphi^t \sigma^1\sigma^3\psi=-i\varphi^t\sigma^2\phi.
\end{equation}

\section{ non-compact Hopf map}\label{sectsu11hopf}

The original (1st) Hopf map is  given by   
\begin{equation}
S^3 \rightarrow S^2\simeq S^3/S^1,
\end{equation}
and its non-compact version may be introduced as  
\begin{equation}
AdS^3\rightarrow H^2\simeq  AdS^3/S^1,
\label{noncompact1sthopf}
\end{equation}
where $AdS^n\simeq SO(n-1,2)/SO(n-1,1)$, and $H^n$ represents an $n$-dimensional two-leaf  hyperboloid that is equivalent to Euclidean $AdS^n\simeq SO(n,1)/SO(n)$.   
$H^2$ with radius $r$ is simply defined as  
\begin{equation}
\eta_{ab}x^ax^b(=x^2+y^2-z^2)=-r^2.
\label{twoleafhyper}
\end{equation}
Apparently, $H^2$ is invariant under the $SO(2,1)$ rotations generated by 
\begin{equation}
J^a=-i\epsilon^{abc}x_b\frac{\partial}{\partial x^c}.
\end{equation}
With a special choice of the vector on the hyperboloid $(x,y,z)=(0,0,\pm r)$, the stabilizer group is found to be the $SO(2)$ rotational group around the $z$-axis, and hence $H^2\simeq SO(2,1)/SO(2)$. 
With polar coordinates, the coordinates on the two-leaf hyperboloid are parameterized as    
\begin{equation}
x=r\sinh \tau \sin\theta,~~y=r\sinh\tau \cos\theta,~~z=\pm r\cosh \tau, 
\label{necparameter}
\end{equation}
where $-\infty<\tau<\infty$ and  $0\le \theta < 2\pi$. 
$z>0$ corresponds to the upper leaf, while $z<0$ does to the lower leaf.
In this paper, we focus on the upper leaf, while the treatment of the lower leaf is completely analogous.

The non-compact Hopf map (\ref{noncompact1sthopf}) is explicitly represented by the mapping from $g$ to $x^a$:
\begin{equation}
gg^{\dagger}=\eta_{ab}x^a \sigma^3 \kappa^b. 
\label{noncompacthopfmatrix}
\end{equation}
Taking the square of both sides and the trace, one may reproduce the hyperboloid constraint  
\begin{equation}
\eta_{ab}x^a x^b=-1,
\end{equation}
where (\ref{su11trans}) and (\ref{sasbtrace}) were used.
(For simplicity, we deal with a hyperboloid with unit radius in the following, unless  otherwise stated.)
With the parameterization of $g$ (\ref{explicitgmatrix}),   $x^a$ are expressed as    
\begin{equation}
x^1=i(u^*v-v^*u),~~x^2=u^*v+v^*u,~~x^3=u^*u+v^*v,
\end{equation}
or, more concisely,
\begin{equation}
\phi\rightarrow x^a=2\phi^{\dagger}\sigma^3 s^a\phi,
\label{noncompHopfphitox}
\end{equation}
where $\phi$ represents the ``non-compact'' Hopf spinor 
\begin{equation}
\phi=
\begin{pmatrix}
u\\
v
\end{pmatrix},
\end{equation}
which satisfies the normalization
\begin{equation}
\phi^{\dagger}\sigma^3\phi=u^*u-v^*v=1.
\end{equation}
From (\ref{noncompHopfphitox}), the hyperboloid condition is readily derived as 
\begin{equation}
\eta_{ab}x^a x^b=-(\phi^{\dagger}\sigma^3\phi)^2=-1.
\end{equation}
With the complex representation $\tilde{s}^a=\frac{1}{2}\tilde{\kappa}^a$, (\ref{noncompHopfphitox}) is rewritten as  
\begin{equation}
\phi\rightarrow x^a=2\phi^t \tilde{s}_a  \sigma^3 \phi^*.
\label{complexnoncomphopfmapspinor}
\end{equation}
Inverting (\ref{noncompHopfphitox}), the non-compact Hopf spinor is expressed as  
\begin{equation}
\phi=\begin{pmatrix}
\sqrt{\frac{1+x^3}{2}}\\
\frac{x^2-ix^1}{\sqrt{2(1+x^3)}}
\end{pmatrix} e^{i\chi}=
\begin{pmatrix}
\cosh \frac{\tau}{2}\\
\sinh \frac{\tau}{2} e^{i\theta}
\end{pmatrix} e^{i\chi},
\end{equation}
where the $U(1)$ phase factor is canceled in the mapping 
(\ref{complexnoncomphopfmapspinor}). 
The non-compact Hopf spinor is equal to the $SU(1,1)$ coherent state formulated in  \cite{barutcommunmathphys21}, which satisfies the coherent state equation 
\begin{equation}
\eta_{ab}x^as^b \phi=-\frac{1}{2}\phi,
\end{equation}
or  
\begin{equation}
\eta_{ab}x^a\phi^t\tilde{s}^b=-\frac{1}{2}\phi^t.
\end{equation}

\subsection{$U(1)$ Connection}

The non-compact Hopf map induces the $U(1)$ connection as 
\begin{equation}
A=\frac{i}{2}tr(g^{\dagger}\sigma^3dg)=i\phi^{\dagger}\sigma^3d\phi,
\end{equation}
which is explicitly evaluated as 
\begin{equation}
A= dx^a A_a = - \frac{I}{2}dx^a
\epsilon_{ab}^{~~3}\frac{x^b}{1+x^3},
\label{gaugefields}
\end{equation}
with $I=1$. In general, $I$ takes an integer, and $I/2$ represents the ``monopole'' charge. 
The corresponding field strengths are given by 
\begin{equation}
F_{ab}=\partial_a A_b -\partial_b A_a=-\frac{I}{2}\epsilon_{abc}x^c.
\end{equation}

\section{Hyperbolic Landau Problem}\label{oneparticleprob}

Here, we explore one-particle quantum mechanics on the surface of a hyperboloid in a monopole background.

\subsection{$SU(1,1)$ Covariant Angular Momenta}

The $SU(1,1)$ covariant angular momenta are given by 
\begin{equation}
\Lambda^a=-i\epsilon^{abc}x_bD_c, 
\end{equation}
where $D_a$ denote covariant derivatives 
\begin{equation}
D_a=\partial_a+iA_a.
\end{equation}
The algebra of the covariant angular momenta is  
\begin{equation}
[\Lambda^a,\Lambda^b]=i\epsilon^{ab}_{~~~c}(\Lambda^c-F^c),
\end{equation}
with $SO(2,1)$ vector field strengths $F^a$  
\begin{equation}
F^{a}=-\frac{1}{2}\epsilon^{abc}F_{bc}=-\frac{I}{2}x^a.
\end{equation}
The covariant angular momenta are tangent to the surface of the hyperboloid, and orthogonal to the field strengths 
\begin{equation}
\eta_{ab}\Lambda^aF^b=\eta_{ab}F^a\Lambda^b=0.
\label{orthogoLamandF}
\end{equation}
The total angular momenta $J^a$ are constructed as  
\begin{equation}
J^a=\Lambda^a+F^a,
\label{totalSU11charge}
\end{equation}
and satisfy the relations
\begin{equation}
[J^a,M^b]=i\epsilon^{ab}_{~~c}M^c,
\label{algebratotalwithotherangmomenta}
\end{equation}
where $M^a=J^a,\Lambda^a$ and $F^a$.  
In particular, when $M^a=J^a$, (\ref{algebratotalwithotherangmomenta}) represents the closed $SU(1,1)$ algebra, and the corresponding $SU(1,1)$ Casimir operator is given by 
\begin{equation}
C=\eta_{ab}J^aJ^b=\eta_{ab}\Lambda^a\Lambda^b-\frac{I^2}{4},
\label{casimirrelation}
\end{equation}
where (\ref{orthogoLamandF}) was used. The eigenvalues of the Casimir operator are 
\begin{equation}
C=-j(j-1),
\label{camilrvalues}
\end{equation}
where, due to the existence of field strengths, $j$ takes   
\begin{equation}
j=-\frac{I}{2}+n+1.
\label{shiftj}
\end{equation}
Here $n$ denotes Landau level (LL) index.

\subsection{One-particle Hamiltonian}

The one-particle Hamiltonian is 
\begin{equation}
H= \frac{1}{2M}\eta_{ab}\Lambda^a\Lambda^b,
\end{equation}
in which the radial kinetic term does not exist, since the particle is confined on the surface of the hyperboloid. With (\ref{casimirrelation}) and (\ref{shiftj}), 
the energy eigenvalues are easily derived as 
\begin{equation}
E_n=\frac{1}{2M}(I(n+\frac{1}{2})-n(n+1)).
\label{su11energyLL}
\end{equation}
Eq.(\ref{su11energyLL}) coincides with the result in Refs.\cite{AP173(1987)185,NPB413(1994)735,amp462005,arXiv:hep-th/0505095,
arXiv:hep-th/0605289,arXiv:hep-th/0605290}.
Unlike the case of the sphere \cite{PRL511983}, the hyperboloid Landau level energy has the maximum 
\begin{equation}
E_{\text{max}}=\frac{I^2}{8M}+\frac{1}{8M}
\end{equation}
at $n=I/2-1/2$.
Meanwhile, the LLL energy is the same in the case of sphere 
\begin{equation}
E_{LLL}=E_{n=0}=\frac{I}{4M}.
\label{bosonicLLLenergy}
\end{equation}
However, the hyperboloid LLL energy is $\it{not}$ the minimum, since (\ref{su11energyLL}) is unbounded as found  at $n\rightarrow \infty$.
By recovering the radius $r$ and taking the thermodynamic limit, $I, r\rightarrow \infty$ with fixed $I/r^2$, Eq.(\ref{su11energyLL}) reproduces the LL energies on the Euclidean plane
\begin{equation}
E_n\rightarrow \omega(n+\frac{1}{2}),
\end{equation}
where $\omega=I/Mr^2$. 

The eigenstates in the LLL are constructed by the symmetric products of the components of the non-compact Hopf spinor
\begin{equation}
u_{m_1,m_2}=\sqrt{\frac{I ! }{m_1 ! m_2 ! }} u^{m_1} v^{m_2},
\label{explicitlllbases}
\end{equation}
where $m_1,m_2 \ge 0$, and $m_1+m_2=I$. 
Since we are concerned with the non-unitary representation, the degeneracy in the LLL becomes finite, and we define the filling fraction as  
\begin{equation}
\nu=N/D \rightarrow 1/m,
\end{equation}
where $N=I+1$ denotes the number of all particles, and   $D=mI+1$ does the number of all states, respectively. The right arrow corresponds to the thermodynamic limit.

\subsection{Coherent State on a Hyperboloid}
With $J^a$ of  $I=1$, the non-compact Hopf spinor satisfies  
\begin{equation}
J^a\phi=-{s^a}\phi,
\end{equation}
and, in the LLL, the $SU(1,1)$ operators are effectively represented as  
\begin{equation}
J^a=-\phi^t \tilde{s}_a\frac{\partial}{\partial\phi},
\end{equation}
where $ \tilde{s}^a={s}_a^t$.
Since $-\tilde{s}_a$ obey the $SU(1,1)$ relations, so do $J^a$. 
The one-particle state aligned with the direction $\Omega^a(\chi)$ on the hyperboloid satisfies the relation 
\begin{equation}
[\Omega^a(\chi)\cdot J_a ] \phi_{\chi}(\phi)
=\frac{I}{2}\phi_{\chi}(\phi),
\end{equation}
and $\phi_{\chi}$ is constructed as 
\begin{equation}
\phi_{\chi}(\phi)
=(\chi^{\dagger}\sigma^3\phi)^{I}=(\alpha^*u-\beta^*v)^
{I},
\end{equation}
where $\chi=(\alpha, \beta)^t$  is related to $\Omega^a(\chi)$ by the relation
\begin{equation}
\Omega^a(\chi)=
{\chi}^{\dagger}\sigma^3\kappa^a\chi.
\label{chiandomegarel}
\end{equation}

\section{Hyperbolic Noncommutative geometry and Hyperbolic Hall Law}\label{sectnoncommutative}

The kinetic term is quenched in LLL, and the LLL limit is realized by simply neglecting $\Lambda^a$. Then, in the limit, from (\ref{totalSU11charge}), one may deduce the relation  
\begin{equation}
x^a\rightarrow X^a= -\alpha L^a,
\end{equation}
with $\alpha=2/I$. While, originally, $x^a$ are the $c$-number coordinates on the hyperboloid,   they are effectively regarded as the $SU(1,1)$ operators in the LLL, and they  satisfy the algebra
\begin{equation}
[X^a,X^b]=-i\alpha\epsilon^{abc}X_c,
\label{fuzzyhyperboloidalgebra}
\end{equation}
which defines the fuzzy hyperboloid \cite{hep-th/0004072,hep-th/0005268}. 
From (\ref{fuzzyhyperboloidalgebra}), the equations of motion are 
 derived as  
\begin{equation}
I^a=\frac{d}{dt}X^a=-i[X^a, V]=-\alpha \epsilon^{abc}x_b E_c, 
\end{equation}
with the electric field $E_a=-\partial_a V$, so one may find the hyperbolic Hall law  
\begin{equation}
\eta_{ab}I^a E^b=0.
\end{equation}

\section{Hyperbolic Quantum Hall Effect}\label{sectqhe}
\subsection{Hyperbolic Laughlin-Haldane Wavefunction}
In the original Haldane's set-up, the Laughlin wavefunction is given by the $SU(2)$ singlet made of the (compact) Hopf spinors 
\cite{PRL511983}, and indeed, such spherical Laughlin-Haldane wavefunction can also be constructed from the stereographic projection from the Laughlin wavefunction on the Euclidean plane. The Laughlin-Haldane wavefunction on a hyperboloid could similarly be derived: we may adopt the $SU(1,1)$ singlet made of the non-compact Hopf spinors  
\begin{equation}
\Phi=\prod_{i<j}(\phi_i^t \sigma^3 R\phi_j)^m=\prod_{i<j}(u_iv_j-v_iu_j)^m,
\label{hyperbolicllin}
\end{equation}
which is consistent with the results in Refs.\cite{NPB413(1994)735,arXiv:hep-th/0505095}. 
The last expression of (\ref{hyperbolicllin}) is superficially equivalent to the 
original Laughlin-Haldane function \cite{PRL511983}, but here, the non-compact Hopf spinors are used. 
Since any two-body state described by the hyperbolic Laughlin-Haldane wavefunction does not have an $SU(1,1)$ angular momentum greater than $m(N-2)$, the hard-core pseudo-potential Hamiltonian is constructed as 
\begin{equation}
H_{h.c.}=\sum_{i<j}\sum_{m(N-2)+1\le J}^{m(N-1)}V_J P_J(i,j),
\end{equation}
where $V_J >0$ denotes the pseudo-potential, and  $P_J$ represents the projection operator to the two-body subspace with the $SU(1,1)$ Casimir index $J$, 
\begin{align}
&P_J(i,j)\nonumber\\
&=\prod_{J'\neq J} \frac{\eta_{ab}(J^a(i)+J^b(i))(J^b(j)+J^b(j)) +J'(J'-{1})}{  J'(J'-1)-J(J-1)}\nonumber\\
&=\prod_{J'\neq J} \frac{2\eta_{ab}J^a(i)J^b(j)-I(\frac{I}{2}-1) +J'(J'-{1})}{J'(J'-1)-J(J-1)}.
\end{align}
In the last equation,  we have used  $\eta_{ab}J^aJ^b=-j(j-1)_{j=-I/2+1}=-{I}/{2}({I}/{2}-1)$.

\subsection{Excitations}

Operators for excitations (quasi-particle and quasi-hole) on a hyperboloid are, respectively, given by   
\begin{subequations}
\begin{align}
&A(\chi)=\prod_i^{N}\chi^{\dagger}R^{\dagger}
\frac{\partial}{\partial\phi_i}=\prod_i^N(\alpha^*\frac{\partial}{\partial v_i}+\beta^*\frac{\partial}{\partial u_i}),\\
&A^{\dagger}(\chi)=\prod_i^{N}\phi^t_i R\sigma^3 \chi
=\prod_i(\alpha v_i -\beta u_i),
\end{align}
\end{subequations}
where $\chi$ specifies the point $\Omega^a(\chi)$ at which excitations are generated, by the relation (\ref{chiandomegarel}).  
Their commutation relations are evaluated as 
\begin{align}
&[A(\chi),A^{\dagger}(\chi)]=1,\nonumber\\
&[A(\chi),A(\chi')]=[A^{\dagger}(\chi),A^{\dagger}(\chi') ]=0,
\end{align}
and $A(\chi)$ and $A^{\dagger}(\chi)$ are interpreted as  
annihilation and creation operators, respectively.
The creation operator satisfies the following commutation relation with angular momentum, 
\begin{equation}
[\Omega_a(\chi) J^a, A^{\dagger}(\chi)]=-\frac{N}{2}A^{\dagger}(\chi).
\label{commuomegajaandaa}
\end{equation}
In particular, at the bottom of the upper leaf $\Omega^a=(0,0,1)$, (\ref{commuomegajaandaa}) becomes 
\begin{equation}
[J^z,A^{\dagger}(\chi)]=\frac{N}{2} A^{\dagger}(\chi),
\end{equation}
which implies that the generation of the quasi-hole pushes each of the particles to the $z$-direction by $1/2$, and the quasi-hole is identified with a charge deficit. 
At  $\nu=1/m$, there are $m$ states per each particle, and the quasi-hole carries the fractional charge $1/m$.

\section{Preliminaries II}\label{sectiosp11group}

For the construction of the spherical SUSY QHE \cite{hep-th/0409230,hep-th/0411137}, the $UOSp(1|2)$ group was used.
The bosonic subgroup of $UOSp(1|2)$ is $SU(2)$, and the graded Hermitian conjugate was adopted to impose a consistent Majorana condition.
Meanwhile,  for the case of the hyperbolic SUSY QHE, we use the $OSp(1|2)$ group
whose subgroup is $SU(1,1)$, and the conventional Hermitian conjugate is adopted \footnote{One may consult Refs.\cite{Hughes1981JMP, dictionaryonsuperalgebras} for detail properties of $UOSp(1|2)$ and $OSp(1|2)$, and their representations.}. 

\subsection{$OSp(1|2)$ Group and Algebra}

Here, we sketch basic structures of the $OSp(1|2)$ group.
The $OSp(1|2)$ group element  $g$ is defined so as to satisfy the relation 
\begin{equation}
g^{\dagger}k g=k,
\end{equation}
and the constraint 
\begin{equation}
\text{sdet} (g)=1.
\label{normalizationSUSYhopfspi}
\end{equation}
Here,
\begin{equation}
k=
\begin{pmatrix}
1 & 0 & 0\\
0 & -1 & 0 \\
0 & 0 & -1
\end{pmatrix},
\label{explicitk}
\end{equation}
and the super-determinant ($\text{sdet}$) is defined in Appendix \ref{defsinsuper}. 
The $g$ is parameterized as  
\begin{equation}
g=
\begin{pmatrix}
u & v^* & \eta^*u+\eta v^*\\
v & u^* & \eta u^*+\eta^* v \\
\eta & -\eta^* & 1-\eta^*\eta
\end{pmatrix},\label{parametrizeospele}
\end{equation}
where $u$ and $v$ are Grassmann even quantities, and $\eta$ is Grassmann odd quantity. 
The inverse of $g$ is $\it{not}$ its simple Hermitian conjugate, but   
\begin{align}
&g^{-1}=  k g^{\dagger}k=
\begin{pmatrix}
u^* & -v^* & -\eta^*\\
-v & u & -\eta \\
-\eta u^*-\eta^* v & \eta^* u+\eta v^* & 1-\eta^*\eta
\end{pmatrix}\nonumber\\
&~~~~~\neq g^{\dagger}=
\begin{pmatrix}
u^* & v^* & \eta^*\\
v & u & -\eta \\
\eta u^*+\eta^* v & \eta^* u+\eta v^* & 1-\eta^*\eta
\end{pmatrix}.
\end{align}
With (\ref{parametrizeospele}), the constraint (\ref{normalizationSUSYhopfspi}) is restated as 
\begin{equation}
u^*u-v^*v-\eta^*\eta=\psi^{\dagger}k\psi=-\psi^t k'\psi^*=1,
\label{normalizationpsi}
\end{equation}
where  $\psi$ denotes the non-compact SUSY Hopf spinor  
\begin{equation}
\psi=
\begin{pmatrix}
u\\
v\\
\eta
\end{pmatrix},
\label{non-compactHopfspinor}
\end{equation}
and 
\begin{equation}
k'=
\begin{pmatrix}
-1 & 0 & 0 \\
0 & 1 & 0 \\
0 & 0 & -1
\end{pmatrix}.
\label{matrixk'}
\end{equation}

The $OSp(1|2)$ algebra is constructed as 
\begin{align}
&[l^a,l^b]=i\epsilon^{ab}_{~~c}l^c,\nonumber\\
&[l^a,l^{\alpha}]=\frac{1}{2}(\kappa^a)_{\beta}^{~~\alpha}l^{\beta},\nonumber\\
&\{l^{\alpha},l^{\beta}\}=\frac{1}{2}(\epsilon^t \kappa_a)^{\alpha\beta}l^a,
\label{ospr12algebra}
\end{align}
where 
\begin{equation}
\epsilon=\epsilon_{\alpha\beta}=
\begin{pmatrix}
0 & 1 \\
-1 & 0 
\end{pmatrix}, ~~\epsilon^t=\epsilon^{\alpha\beta}=
\begin{pmatrix}
0 & -1 \\
1 & 0 
\end{pmatrix}.
\end{equation}
The $OSp(1|2)$ Casimir operator is  given by 
\begin{equation}
C=\eta_{ab}l^a l^b-\epsilon_{\alpha\beta}l^{\alpha}l^{\beta},
\end{equation}
and its eigenvalues are 
\begin{equation}
C=-j(j-\frac{1}{2}),
\end{equation}
with $j=1/2,1,{3}/2,2,\cdots$. It is noted the Casimir index begins from $1/2$ not $0$.
The fundamental representation of the $OSp(1|2)$ algebra is  
\begin{equation}
l^a=\frac{1}{2}
\begin{pmatrix}
\kappa^a & 0 \\
0 & 0
\end{pmatrix},
~~l^{\alpha}=
\frac{1}{2}
\begin{pmatrix}
0 & \tau^{\alpha}\\
-(\epsilon \tau^{\alpha})^t & 0 
\end{pmatrix},
\label{originalospfundmatr}
\end{equation}
and is normalized as  
\begin{equation}
\text{str} (l^a l^b)=-\frac{1}{2}\eta^{ab},~~~\text{str} (l^{\alpha}l^{\beta})=\frac{1}{2}\epsilon^{\alpha\beta},~~
\text{str} (l^a l^{\alpha})=0,
\label{normalizationcondlanal}
\end{equation}
where the super-trace ($\text{str}$) is defined in Appendix \ref{defsinsuper}.
When $l^a$ and $l^{\alpha}$ satisfy the $OSp(1|2)$ algebra, 
\begin{equation}
-l_a, ~~~ d^{\alpha}\equiv (\sigma^1)_{\beta}^{~~\alpha}(l^{\beta})^t
\label{anotherosprep}
\end{equation}
also satisfy the algebra.  $-l_a$ and $d^{\alpha}$ are related to $l^a$ and $l^{\alpha}$ as  
\begin{equation}
-l_a=k l^a k,~~d^{\alpha}=k l^{\alpha} k,
\end{equation}
with $k$ (\ref{explicitk}).

\subsection{Complex Representation}\label{complexOSp(1|2)}

The complex representation of (\ref{originalospfundmatr}) is constructed as 
\begin{equation}
\tilde{l}^a=-{l^a}^*,
~~\tilde{l}^{\alpha}= \epsilon_{\alpha\beta}d^{\beta},
\label{complexrepreo}
\end{equation}
and related to $l^a$ and $l^{\alpha}$ by the unitary transformation 
\begin{equation}
\tilde{l}^a=\mathcal{R}^{\dagger}l^a\mathcal{R},~~\tilde{l}^{\alpha}=
\mathcal{R}^{\dagger}l^{\alpha}\mathcal{R},
\end{equation}
where  
\begin{equation}
\mathcal{R}=
\begin{pmatrix}
0 & 1 & 0 \\
1 & 0 & 0 \\
0 & 0 & 1
\end{pmatrix}.
\end{equation}
The properties of $\mathcal{R}$ are summarized as 
\begin{align}
&\mathcal{R}=\mathcal{R}^t=\mathcal{R}^{\dagger}=\mathcal{R}^{-1},\\
&\mathcal{R}^2=(\mathcal{R}^t)^2=
\begin{pmatrix}
1 & 0 & 0 \\
0 & 1 & 0 \\
0 & 0 & 1
\end{pmatrix}.
\end{align}
Then, the charge conjugation of $\psi$ is determined as  
\begin{equation}
\psi_c=\mathcal{R}^{\dagger}\psi^*,
\end{equation}
and, without using complex conjugation, the $OSp(1|2)$ singlet can be constructed as   
\begin{equation}
(\psi_c)^{\dagger}k\psi'=\psi^t \mathcal{R}k \psi'=-(uv'-vu'+\eta\eta').
\label{Ospinv1} 
\end{equation}

For later convenience, we introduce another complex representation 
\begin{equation}
{j}^a=l^*_a,~~{j}^{\alpha}=\epsilon_{\alpha\beta}l^{\beta}
\label{defofjs}
\end{equation}
whose original representation is  $-l_a$ and $d^{\alpha}$. 
Eq.(\ref{anotherosprep}) and Eq.(\ref{defofjs}) are related by the unitary transformation 
\begin{equation}
j^a=\mathcal{R}^{\dagger}(-l_a)\mathcal{R},~~j^{\alpha}=
\mathcal{R}^{\dagger}d^{\alpha}\mathcal{R}.
\end{equation}
It should be noticed that $j^a$ and $j^{\alpha}$ are linearly dependent on $l^a$ and $l^{\alpha}$, while $\tilde{l}^a$ and $\tilde{l}^{\alpha}$ are not, because $\tilde{l}^{\alpha}=(-1)^{\alpha+1}(l^{\alpha})^t$ cannot be expressed by linear combinations of $l^a$ and $l^{\alpha}$. In the following, $j^a$ and $j^{\alpha}$ will be used rather than $\tilde{l}^a$ and $\tilde{l}^{\alpha}$.    
While (\ref{Ospinv1}) is not invariant under the $OSp(1|2)$ transformation generated by $j^a$ and $j^{\alpha}$,  
\begin{equation}
\psi^tk\mathcal{R}\psi'=uv'-vu'-\eta\eta'
\end{equation}
is invariant.
The two complex representations (\ref{complexrepreo}) and (\ref{defofjs}) are simply related as
\begin{equation}
j^a=k\tilde{l}^ak,~~~j^{\alpha}=k\tilde{l}^{\alpha}k.
\end{equation}
Further, they are related to $l^a$ and $l^{\alpha}$ by the unitary transformation
\begin{equation}
{j}^a=\mathcal{K}^t l^a \mathcal{K},
~~{j}^{\alpha}=\mathcal{K}^t l^{\alpha}\mathcal{K}
\label{jajalphamatrix}
\end{equation}
where 
\begin{equation}
\mathcal{K}=\mathcal{R} k=k'\mathcal{R}=
\begin{pmatrix}
0 & -1  & 0 \\
1 & 0 & 0 \\
 0 & 0 & -1
\end{pmatrix}.
\end{equation}
The properties of $\mathcal{K}$ are similar to those of $\mathcal{R}$: 
\begin{equation}
\mathcal{K}^t=\mathcal{K}^{\dagger}=\mathcal{K}^{-1},
\end{equation}
but $\mathcal{K}\neq \mathcal{K}^t$, and 
\begin{equation}
\mathcal{K}^2=(\mathcal{K}^t)^2=
\begin{pmatrix}
-1 & 0 & 0 \\
0 & -1 & 0 \\
0 & 0 & 1
\end{pmatrix}
=k k'=k'k.
\end{equation}
$k$ and $k'$ are constructed from the products of  $\mathcal{K}$ and $\mathcal{R}$ as 
\begin{equation}
k=\mathcal{R}\mathcal{K},~~~k'=\mathcal{K}\mathcal{R}, 
\end{equation}
and  related as 
\begin{equation}
k=\mathcal{R}^t k'\mathcal{R}=\mathcal{K}^t k'\mathcal{K},~~~
k'=\mathcal{R}^t k\mathcal{R}=\mathcal{K}^t k  \mathcal{K}.
\end{equation}

\section{The Non-compact SUSY Hopf Map}\label{sectsusyhopfmap}

The (original) SUSY Hopf map  
\begin{equation} 
S^{3|2}\rightarrow S^{2|2}\simeq S^{3|2}/S^1
\end{equation}
was introduced in Ref.\cite{PLB193p61}, and the accompanying bundle structure has been  well examined in Refs.\cite{{J.Math.Phys.31(1990),math-ph/9907020}} and Ref.\cite{hep-th/0409230}. 
Here, we explore the non-compact version of it
\begin{equation}
AdS^{3|2}\rightarrow H^{2|2}\simeq AdS^{3|2}/S^1,
\end{equation}
where the super-hyperboloid $H^{2|2}$ or Euclidean $AdS^{2|2}$ is defined so as to satisfy the condition
\begin{equation}
\eta_{ab}x^a x^b-\epsilon_{\alpha\beta}\theta^{\alpha}\theta^{\beta}=-1.
\label{superhypercond}
\end{equation}
Apparently, the condition is invariant under the $OSp(1|2)$ transformations generated by 
\begin{align}
&L^a=-i\epsilon^{abc}x_b\partial_c+\frac{1}{2}(\kappa^a)_{\beta}^{~~\alpha}\theta^{\beta}\partial_{\alpha},\nonumber\\
&L^{\alpha}=\frac{1}{2}(\epsilon^t \kappa^a)^{\alpha\beta}x_a\partial_{\beta}-\frac{1}{2}(\kappa^a)_{\beta}^{~~\alpha}\theta^{\beta}\partial_a,
\end{align}
and $H^{2|2}$ manifestly possesses the $OSp(1|2)$ symmetry.
The non-compact SUSY Hopf map is explicitly constructed as 
\begin{equation}
g\rightarrow g k^3 g^{\dagger}=\delta_{ab}x^a k^b +(\sigma^1)_{\alpha\beta}\theta^{\alpha}k^{\beta}, \label{matrixnoncompacthopf}
\end{equation}
where $k^a=k l^a$ and $k^{\alpha}=k l^{\alpha}$ are     
\begin{align}
&k^1=\frac{1}{2}
\begin{pmatrix}
0 & i & 0 \\
-i & 0 & 0 \\
0 & 0 & 0
\end{pmatrix},~~
k^2=\frac{1}{2}
\begin{pmatrix}
0 & 1 & 0 \\
1 & 0 & 0 \\
0 & 0 & 0
\end{pmatrix},~~k^3=\frac{1}{2}
\begin{pmatrix}
1 & 0 & 0 \\
0 & 1 & 0 \\
0 & 0 & 0
\end{pmatrix},\nonumber\\
&k^{\theta_1}=\frac{1}{2}
\begin{pmatrix}
0 & 0 & 1 \\
0 & 0 & 0 \\
0 & -1 & 0
\end{pmatrix},~~
k^{\theta_2}=\frac{1}{2}
\begin{pmatrix}
0 & 0 & 0 \\
0 & 0 & -1 \\
1 & 0 & 0
\end{pmatrix}.
\end{align}
Though $k^a$ and $k^{\alpha}$ are ``Hermitian'' in the sense that 
\begin{equation}
{k^a}^{\dagger}=k^a,~~{k^{\alpha}}^{\dagger}=(\sigma^1)_{\beta}^{~~\alpha}k^{\beta},
\end{equation}
they do not form a closed algebra.
With the normalization (\ref{normalizationcondlanal}), it is not difficult to see that $x^a$ and $\theta^{\alpha}$ (introduced by (\ref{matrixnoncompacthopf})) indeed 
 satisfy the super-hyperboloid condition (\ref{superhypercond}).
With (\ref{parametrizeospele}), $x^a$ and $\theta^{\alpha}$ are expressed as  
\begin{align}
&x^1=i(u^*v-v^*u),~~x^2=u^*v+v^*u,~~x^3=u^*u+v^*v,\nonumber\\
&\theta^1=u^*\eta-\eta^* v,~~\theta^2=\eta^*u-\eta v^*,
\end{align}
or, more compactly,   
\begin{equation}
x^a=2\psi^{\dagger} k^a\psi,~~~\theta^{\alpha}=2\psi^{\dagger} k^{\alpha}\psi,
\label{SUSYHopfmapexpli}
\end{equation}
where $\psi$ is the non-compact SUSY Hopf spinor (\ref{non-compactHopfspinor}).
From the ``Hermiticity'' of $k^a$ and $k^{\alpha}$, $x^a$ and $\theta^{\alpha}$ are ``real'' in the sense that 
\begin{equation}
{x^a}^*=x^a,~~{\theta^{\alpha}}^*=(\sigma^1)_{\beta}^{~~\alpha}\theta^{\beta}.
\end{equation}
Namely, $\theta=(\theta^1,\theta^2)^t$ is an $SO(2,1)$ Majorana-spinor. 
From the non-compact SUSY Hopf map (\ref{SUSYHopfmapexpli}) and the constraint (\ref{normalizationpsi}), it is readily confirmed that $x^a$ and $\theta^{\alpha}$ satisfy the  condition (\ref{superhypercond}), since   
\begin{equation}
\eta_{ab}x^ax^b-\epsilon_{\alpha\beta}\theta^{\alpha}\theta^{\beta}
=-(\psi^{\dagger}k\psi)^2=-1.
\end{equation}

With the complex representation, the non-compact SUSY Hopf map (\ref{SUSYHopfmapexpli}) is restated as 
\begin{equation}
x^a=2\psi^t  k'^a \psi^*,~~~\theta^{\alpha}=2\psi^t k'^{\alpha}\psi^*,
\label{complexSUSYhopfmap}
\end{equation}
where 
\begin{equation}
k'^a\equiv j^ak'=-\eta_{ab}k^b  ,~~~k'^{\alpha}\equiv j^{\alpha}k'=(\sigma^1)_{\beta}^{~~\alpha}k^{\beta}.
\end{equation}

Inverting (\ref{SUSYHopfmapexpli}), the non-compact SUSY Hopf spinor is expressed by $x^a$ and $\theta^{\alpha}$, up to the $U(1)$ phase factor, as 
\begin{equation}
\psi=\frac{1}{\sqrt{2(1+x^3)}}
\begin{pmatrix}
({1+x^3})( 1-\frac{1}{4(1+x^3)}\theta\epsilon\theta )\\
({x^2-ix^1})(1+\frac{1}{4(1+x^3)}\theta\epsilon\theta)\\
(1+x^3)\theta^1+(x^2-ix^1)\theta^2
\end{pmatrix}\cdot e^{i\chi},
\label{explicitnonsusyhopfspinor}
\end{equation}
which satisfies the supercoherent equation 
\begin{equation}
\eta_{ab}l^a\psi x^b-\epsilon_{\alpha\beta}l^{\alpha}\psi \theta^{\beta}=-\frac{1}{2}\psi,
\end{equation}
or, in the complex representation,
\begin{equation}
\eta_{ab}x^a\psi^t j^b -\epsilon_{\alpha\beta}\theta^{\alpha}\psi^t {j}^{\beta}=\frac{1}{2}\psi^t.
\end{equation}
Thus, the non-compact SUSY Hopf spinor is equivalent to the $OSp(1|2)$ supercoherent state in Ref.\cite{JMP32(1991)3381}.

\subsection{$U(1)$ connection}
The non-compact SUSY Hopf map (\ref{matrixnoncompacthopf}) or (\ref{SUSYHopfmapexpli}) is invariant under the $U(1)$ gauge transformation:  
\begin{equation}
g\rightarrow g e^{2i\alpha l^3}
\label{U1transofg}
\end{equation}
or 
\begin{equation}
\psi\rightarrow e^{i\alpha}\psi.
\end{equation}
Such gauge freedom induces a $U(1)$ connection on a super-hyperboloid as   
\begin{equation}
A=i\text{str} (k^3 g^{\dagger}k dg)=i\psi^{\dagger}k d\psi.
\end{equation}
Accompanied with the $U(1)$ gauge transformation (\ref{U1transofg}), $A$ is transformed as 
\begin{equation}
A\rightarrow A+d\alpha,
\end{equation}
as expected.
With the explicit form of the non-compact SUSY Hopf spinor (\ref{explicitnonsusyhopfspinor}), the components of the $U(1)$ gauge field 
\begin{equation}
A=dx^a A_a +d\theta^{\alpha}A_{\alpha}
\end{equation}
are evaluated as 
\begin{align}
&A_a=-\frac{I}{2}\epsilon_{ab}^{~~3}\frac{x^b}{1+x^3}\biggl(1+\frac{2+x^3}{2(1+x^3)}\theta\epsilon\theta\biggr),\nonumber\\
&A_{\alpha}=-i\frac{I}{2}x^a (\theta \kappa_a\epsilon)_{\alpha},
\end{align}
with $I=1$. $I/2$ represents the ``supermonopole'' charge with integer $I$. 
Their complex conjugations are given by  
\begin{equation}
A_a^*=A_a,~~A_{\alpha}^*=-(\sigma^1)_{\alpha}^{~~\beta}A_{\beta}.
\end{equation}
The super field strengths  
\begin{align}
&F_{ab}=\partial_a A_b- \partial_b A_a,\nonumber\\
&F_{a\alpha}=\partial_a A_{\alpha}-\partial_{\alpha}A_a,\nonumber\\
&F_{\alpha\beta}=\partial_{\alpha}A_{\beta}+\partial_{\beta}A_{\alpha},
\end{align}
are also evaluated as 
\begin{align}
&F_{ab}=   -\frac{I}{2}\epsilon_{abc}x^c(1+\frac{3}{2}\theta\epsilon\theta), \nonumber\\
&F_{a\alpha}=-i\frac{I}{2}(\theta\kappa_b \epsilon)_{\alpha}
(\delta^{b}_{~a}-3x_ax^b), \nonumber\\
&F_{\alpha\beta}=-iI(\kappa_a \epsilon)_{\alpha\beta}x^a (1+
\frac{3}{2}\theta\epsilon\theta).
\end{align}

\section{hyperbolic SUSY Landau  Problem}\label{sectsusyoneparticle}

The Landau problem is inspected on the surface of a super-hyperboloid in the supermonopole background. 
\subsection{$OSp(1|2)$ Covariant Angular Momenta}

There are two-kinds of  covariant angular momenta: one is bosonic and the other is fermionic, 
\begin{align}
&\Lambda^a=-i\epsilon^{abc}x_b D_c+\frac{1}{2}(\kappa^a)_{\beta}^{~~\alpha}\theta^{\beta} D_{\alpha},\nonumber\\
&\Lambda^{\alpha}=\frac{1}{2}(\epsilon^t \kappa^a)^{\alpha\beta}x_a D_{\beta}-\frac{1}{2}(\kappa^a)_{\beta}^{~~\alpha}\theta^{\beta}D_a, 
\end{align}
where the covariant derivatives are defined by  
\begin{equation}
D_a=\partial_a +iA_a,~~~D_{\alpha}=\partial_{\alpha}+iA_{\alpha}.
\end{equation}
The covariant angular momenta satisfy the relations 
\begin{align}
&[\Lambda^a,\Lambda^b]=i\epsilon^{ab}_{~~c}(\Lambda^c-F^c),\nonumber\\
&[\Lambda^a,\Lambda^{\alpha}]=\frac{1}{2}(\kappa^a)_{\beta}^{~~\alpha}(\Lambda^{\beta}-F^{\beta}),\nonumber\\
&\{\Lambda^{\alpha},\Lambda^{\beta}\}=\frac{1}{2}(\epsilon^t \kappa_a)^{\alpha\beta}(\Lambda^a-F^a),
\end{align}
where 
\begin{equation}
F^a=-\frac{I}{2}x^a,~~F^{\alpha}=-\frac{I}{2}\theta^{\alpha}, 
\end{equation}
which are the angular momenta of the supermonopole gauge fields, and are orthogonal to the covariant angular momenta  
\begin{equation}
\eta_{ab}\Lambda^a F^b-\epsilon_{\alpha\beta}\Lambda^{\alpha}F^{\beta}=\eta_{ab}F^a\Lambda^b -\epsilon_{\alpha\beta}F^{\alpha}\Lambda^{\beta}
=0.
\label{orthogonalitylambdafsusy}
\end{equation}
The conserved SUSY angular momenta are constructed as 
\begin{equation}
J^a=\Lambda^a+F^a,~~J^{\alpha}=\Lambda^{\alpha}+F^{\alpha},
\label{totalconserveanglsusy}
\end{equation}
and they generate the $OSp(1|2)$ transformations  
\begin{align}
&[J^a,M^b]=i\epsilon^{ab}_{~~~c}M^c,\nonumber\\
&[J^a,M^{\alpha}]=\frac{1}{2}(\kappa^a)_{\beta}^{~~\alpha}M^{\beta},\nonumber\\
&\{J^{\alpha},M^{\beta}\}=\frac{1}{2}(\epsilon^t \kappa_a)^{\alpha\beta}M^a,
\end{align}
where $M^a=J^a, \Lambda^a, F^a$ and $M^{\alpha}= J^{\alpha}, 
\Lambda^{\alpha}, F^{\alpha}$.
The corresponding $OSp(1|2)$ Casimir operator is given by  
\begin{equation}
\eta^{ab}J^a J^b-\epsilon_{\alpha\beta}J^{\alpha}J^{\beta}=
\eta^{ab}\Lambda^a \Lambda^b-\epsilon_{\alpha\beta}\Lambda^{\alpha}\Lambda^{\beta} -\frac{I^2}{4},
\label{superjsquare}
\end{equation}
where (\ref{orthogonalitylambdafsusy}) and  
\begin{equation}
\eta_{ab}F^a F^b-\epsilon_{\alpha\beta}F^{\alpha}F^{\beta}=-\frac{I^2}{4}
\end{equation}
were used. 
The Casimir operator takes the eigenvalues 
\begin{equation}
\eta_{ab}J^aJ^b-\epsilon_{\alpha\beta}J^{\alpha}J^{\beta}=-j(j-\frac{1}{2})
\label{eigenvaluessuperjsquare}
\end{equation}
with 
\begin{equation}
j=-\frac{I}{2}+n+\frac{1}{2}.
\end{equation}
Here, $n$ denotes the super LL index.

\subsection{One-particle Hamiltonian}

The one-particle Hamiltonian is given by 
\begin{equation}
H=\frac{1}{2M}(\eta_{ab}\Lambda^a\Lambda^b-\epsilon_{\alpha\beta}\Lambda^{\alpha}\Lambda^{\beta}),\label{superoneparticleham}
\end{equation}
and is a supersymmetric Hamiltonian in the sense that it is invariant under the $OSp(1|2)$ transformation.
From (\ref{superjsquare}) and (\ref{eigenvaluessuperjsquare}), its energy eigenvalues are derived as  
\begin{equation}
E_n=\frac{1}{2M}(I(n+\frac{1}{4})-n(n+\frac{1}{2})).
\end{equation}
The energy takes the maximum 
\begin{equation}
E_{\text{max}}=\frac{I^2}{8M}+\frac{1}{32M}
\end{equation}
at $n=I/2-1/4$, and the LLL energy is 
\begin{equation}
E_{LLL}=E_{n=0}=\frac{I}{8M},
\end{equation}
which is equal to the LLL energy on a supersphere \cite{hep-th/0409230}, and is also equal to the half of the LLL energy in the original hyperbolic case (\ref{bosonicLLLenergy}). 
Just as in the original hyperboloid case, the energy eigenvalues on a super-hyperboloid have the maximum, but are unbounded from below.
Since we are concerned with the non-unitary representation of the $OSp(1|2)$ group, 
the degeneracy in the LLL becomes finite and the LLL bases are constructed from the symmetric products of the components of the non-compact SUSY Hopf spinor as 
\begin{align}
&u_{m_1m_2}=\sqrt{\frac{I!}{m_1!m_2!}}u^{m_1}v^{m_2},\nonumber\\
&\eta_{n_1n_2}=\sqrt{\frac{I!}{n_1 ! n_2 !}}u^{n_1}v^{n_2}\eta,
\label{explicitsusylllbases}
\end{align}
where $m_1, m_2, n_1, n_2\ge 0$, and $m_1+m_2=n_1+n_2+1=I$. The degeneracy in the LLL is explicitly given by 
\begin{equation}
D=(I+1)+I=2I+1,
\end{equation}
and thus, the super LLL is almost doubly degenerate compared to the original (bosonic) case.
The filling fraction is usually defined by $N/D$, where $D$ denotes the total number of states $D=D_B+D_F$ ($D_{B}$ and  $D_{F}$ are the total numbers of bosonic and fermionic states, respectively), but  for later convenience, we define the filling fraction  as in the original (bosonic) case 
\begin{equation}
\nu=N/D_B=I/(mI+1)\rightarrow 1/m,
\end{equation}
where the right arrow represents the thermodynamic limit.

\subsection{Supercoherent State on a Super-hyperboloid}

The non-compact SUSY Hopf spinor is equivalent to the supermonopole harmonics with the minimum monopole charge $I/2=1/2$: 
\begin{equation}
J^a_{(I=1)}\psi=(j^a)^t\psi,~~~J^{\alpha}_{(I=1)}\psi=(j^{\alpha})^t\psi.
\end{equation}
where  $j^a$ and $j^{\alpha}$ are given by (\ref{defofjs}).  Therefore, 
in the LLL, $J^a$ and $J^{\alpha}$ are effectively represented as 
\begin{equation}
J^a=\psi^t {j^a}\frac{\partial}{\partial\psi},~~
J^{\alpha}=\psi^t {j}^{\alpha}\frac{\partial}{\partial\psi}.
\label{effectivejaandjalpha}
\end{equation}
The one-particle state aligned with the direction 
$(\Omega^a,\Omega^{\alpha})$, 
\begin{equation}
\Omega^a(\chi)=2\chi^{\dagger}k^a\chi,~~\Omega^{\alpha}(\chi)=2\chi^{\dagger}k^{\alpha}\chi,
\label{relchiomegasusy}
\end{equation}
is represented as  
\begin{equation}
\psi_{\chi}(\psi)
=(\chi^{\dagger}k\psi)^{I}
=(\alpha^*u-\beta^*v-\xi^*\eta)^{I}.
\end{equation}
Indeed, $\psi_{\chi}$ satisfies the equation
\begin{equation}
[\eta_{ab}\Omega^a(\chi){J^{b}}-\epsilon_{\alpha\beta}\Omega^{\alpha}(\chi){J^{\beta}}]\psi_{\chi}(\psi)
=\frac{I}{2}\psi_{\chi}(\psi).
\end{equation}

\section{Hyperbolic Super Fuzzy geometry and Hyperbolic Super Hall Law}\label{susynoncommutativegeo}

Based on similar discussions developed in Sec.\ref{sectnoncommutative}, one may deduce the non-commutative relation on a super-hyperboloid.
From the relation (\ref{totalconserveanglsusy}), in the LLL limit 
($\Lambda^a,~\Lambda^{\alpha} \rightarrow 0$), the coordinates on 
a super-hyperboloid are  regarded as the $OSp(1|2)$ operators
\begin{equation}
x^a\rightarrow X^a= -\alpha L^a,~~~\theta^{\alpha}\rightarrow \Theta^{\alpha}=-\alpha L^{\alpha},
\end{equation}
which satisfy the fuzzy super-algebra
\begin{align}
&[X^a,X^b]=-i\alpha\epsilon^{abc}X_c,\nonumber\\
&[X^a,\Theta^{\alpha}]=-i\frac{\alpha}{2}(\kappa^a)_{\beta}^{~\alpha}\Theta^{\beta},\nonumber\\
&\{\Theta^{\alpha},\Theta^{\beta}\}=-\frac{\alpha}{2}(\epsilon^t\kappa^a)^{\alpha\beta}X_a,
\label{fuzzysuperhyperboloidalgebras}
\end{align}
where $\alpha=2R/I$.
The super-algebra (\ref{fuzzysuperhyperboloidalgebras}) defines a fuzzy supermanifold that could be called the fuzzy super-hyperboloid \footnote{To the author's knowledge, fuzzy super-hyperboloids have not explicitly appeared in the literature, but they might be classical solutions of the supermatrix model, like fuzzy superspheres \cite{hep-th/0311005}.}.
From (\ref{fuzzysuperhyperboloidalgebras}), the super Hall currents are derived as  
\begin{align}
&I^a=\frac{d}{dt}X^a=-i[X^a, V]\nonumber\\
&~~~=-\alpha \epsilon^{abc}x_b E_c+i\frac{\alpha}{2}(\kappa^a)_{\alpha}^{~~\beta}\theta^{\alpha}E_{\beta},\nonumber\\
&I^{\alpha}=\frac{d}{dt}\Theta^{\alpha}=-i[\Theta^{\alpha},V] \nonumber\\
&~~~=i\frac{\alpha}{2}(\epsilon^t\kappa^a)^{\alpha\beta}x_a E_{\beta}+i\frac{\alpha}{2}(\kappa^a)_{\beta}^{~~\alpha}\theta^{\beta}E_a,
\label{hyperbolicsuperhallcurrents}
\end{align}
where  $E_a=-\partial_a V$ and $E_{\alpha}=\partial_{\alpha} V$, and the super-hyperbolic version of  Hall law is confirmed as    
\begin{equation}
\eta_{ab}I^a E^b-\epsilon_{\alpha\beta}I^{\alpha}E^{\beta}=0.
\end{equation}

\section{hyperbolic SUSY Quantum Hall Effect}\label{sectsusyqhe}

\subsection{Hyperbolic SUSY Laughlin-Haldane Wavefunction}

It may be natural to adopt $OSp(1|2)$ singlet function as a  hyperbolic SUSY Laughlin-Haldane  wavefunction 
\begin{equation}
\Psi =  \prod_{i<j}^N(\psi^t_i k \mathcal{R} \psi_j )^m =\prod_{i<j}(u_iv_j-v_iu_j-\eta_i\eta_j)^m. 
\label{superhyperLlin}
\end{equation}
Indeed, (\ref{superhyperLlin}) is invariant under the $OSp(1|2)$ transformations generated by (\ref{effectivejaandjalpha}), and superficially takes the same form of the spherical SUSY Laughlin-Haldane wavefunction proposed in Ref.\cite{hep-th/0411137}, but the non-compact SUSY Hopf spinors are used as here.
The corresponding hard-core pseudo-potential Hamiltonian is constructed as 
\begin{equation}
H_{h.c.}=\sum_{i<j}\sum_{m(N-2)+1/2\le J}^{m(N-1)}V_JP_J(i,j).
\end{equation}
Here, $P_J$ is the projection operator of the two-body subspace of the $OSp(1|2)$ index $J$: 
\begin{align}
&P_J(i,j)\nonumber\\
&\!\!\!=\!\!\!\prod_{J'\neq J} \!\!\frac{2\eta_{ab}J^a(i)J^b(j)\!-\!\epsilon_{\alpha\beta}J^{\alpha}(i)J^{\beta}(j)-\frac{I}{2}(I\!-\!1) \!+\!J'(J'-\frac{1}{2})}{ J'(J'-\frac{1}{2})-J(J\!-\!\frac{1}{2})  },
\end{align}
where we have used  $\eta_{ab}J^aJ^b-\epsilon_{\alpha\beta}J^{\alpha}J^{\beta}=-j(j-{1}/{2})_{j=-I/2+1/2}=-{I}/{2}({I}/{2}-{1}/{2})$.
The hyperbolic SUSY Laughlin-Haldane wavefunction is rewritten as 
\begin{equation}
\Psi=\exp\biggl(-m\sum_{i<j}^N\frac{\eta_i\eta_j}{u_iv_j-v_iu_j}\biggr)\cdot\Phi,
\end{equation}
where $\Phi$ is the original hyperbolic Laughlin-Haldane wavefunction (\ref{hyperbolicllin}).
Expanding the exponential, we obtain   
\begin{align}
&\Psi=\Phi-m\sum_{i<j}\frac{\eta_i\eta_j}{u_iv_j-v_iu_j}\cdot\Phi\nonumber\\
&~~~~~~~+\frac{1}{2}
\biggl(m\sum_{i<j}\frac{\eta_i\eta_j}{u_iv_j-v_iu_j}\biggr)^2\cdot\Phi+\cdots \nonumber\\
&~~~~~~~+
\eta_i\eta_2\cdots \eta_N (-m)^{N/2}  Pf  \biggl(\frac{1}{u_iv_j-v_iu_j}\biggr)\cdot \Phi.
\end{align}
One may find that both the original Laughlin and the Moore-Read Pfaffian wavefunctions appear in the expansion: the former appears as the first term, and the latter as the last term.
Thus, the two quantum Hall wavefunctions are ``unified'' in the SUSY formalism.

\subsection{Excitations}

Operators for excitations (quasi-particle and quasi-hole) on a super-hyperboloid are, respectively, constructed as  
\begin{align}
&A(\chi)= \prod_i \chi^{\dagger} \mathcal{R}\frac{\partial}{\partial\psi_i} = \prod_i \chi^{\dagger}  \mathcal{K}k \frac{\partial}{\partial\psi_i}\nonumber\\
&~~~~~~=\prod_i ( \alpha^*\frac{\partial}{\partial v_i}   +\beta^*\frac{\partial}{\partial u_i}  
+ \xi^*\frac{\partial}{\partial\eta_i}),\nonumber\\
&A^{\dagger}(\chi)= \prod_i  \psi^t_i \mathcal{K} \chi =\prod_i \psi^t_i \mathcal{R}k \chi\nonumber\\
&~~~~~~~=\prod_{i}(\alpha v_i -\beta u_i +\xi\eta_i),
\end{align}
where $\chi$ specifies the point on a super-hyperboloid by the relation (\ref{relchiomegasusy}). 
Their commutation relations are derived as  
\begin{align}
&[A(\chi),A^{\dagger}(\chi)]=1,\nonumber\\
&[A(\chi),A(\chi')]=[A^{\dagger}(\chi),A^{\dagger}(\chi')]=0,
\end{align}
which imply that $A(\chi)$ and $A^{\dagger} (\chi)$ are interpreted as annihilation and creation operators, respectively.
The angular momentum of the quasi-hole follows from 
\begin{equation}
[\Omega_a(\chi) J^a-\epsilon_{\alpha\beta}\Omega^{\alpha}(\chi)J^{\beta}, A^{\dagger}(\chi)]=-\frac{N}{2}A^{\dagger}(\chi),
\label{angulerquasisusy}
\end{equation}
which suggests that the excitation carries the fractional charge $1/m$, in the SUSY QHE also.

\section{Summary and Discussion}\label{summarysection}

Based on the non-compact version of the SUSY Hopf map, we developed a formulation of the QHE on a super-hyperboloid, where the conventional definitions of Hermitian and complex conjugations were used, unlike for the spherical SUSY QHE.
Using $OSp(1|2)$ group theoretical methods, we derived super Landau level energies and the  non-unitary representation of LLL bases.
The Landau level on a super-hyperboloid has the maximum energy, while LLL energy is equivalent to that on a supersphere.
We constructed the Laughlin wavefunction, the hard-core pseudo-potential Hamiltonian and fractionally charged excitations on a super-hyperboloid. 
The hyperbolic SUSY Laughlin-Haldane wavefunction superficially takes the same form as in the spherical QHE, but the non-compact 
Hopf spinors were used in the present formalism. In the LLL, the hyperbolic fuzzy super-geometry naturally emerges. 
It was confirmed that the particular properties in the original hyperbolic QHE were observed in the hyperbolic SUSY QHE. 

There might be many directions to be pursued from the present model.
One apparent direction is to explore extensions of the QHE on other 
non-compact manifolds. 
In particular, the exploration of a non-compact QHE with $SO(3,2)$ symmetry would be interesting, since it is a natural non-compact version of the four-dimensional QHE.
As close analogies between twistor and QHE have been pointed out in Refs.\cite{cond-mat/0211679,cond-mat/0401224}, in the LLL of the model, the $SO(3,2)$ symmetry will naturally be enhanced to $SU(2,2)$ conformal symmetry. 
Then, the $SO(3,2)$ version of  noncompact QHE appears to realize a more direct relationship to twistor theory. 
The study of topological order of the SUSY QHE is another intriguing topic.   
Since the SUSY gives a unified picture of quantum liquids with different topological orders, $i.e.$, Laughlin and Moore-Read states, analyses of the topological order in the SUSY QHE could be important in understanding ``transitions'' between such topologically different quantum liquids. 
We hope the hyperbolic SUSY QHE will be a starting point for such stimulating future directions.

\section*{ACKNOWLEDGMENTS}

I would like to thank Professor Joris Van der Jeugt for telling me about Ref.\cite{Hughes1981JMP} and for useful conversations about irreducible representations of $SU(1,1)$ and $OSp(1|2)$ groups at the XXVII international colloquium ``Group Theoretical Methods in Physics'' (GROUP 27).

\appendix

\section{Several definitions in supergroup}\label{defsinsuper}

When a supermatrix is given by the form 
\begin{equation}
M=
\begin{pmatrix}
A & B \\
C & D
\end{pmatrix}
\end{equation}
($A$ and $D$ are Grassmann-even block components, and $B$ and $C$ are Grassmann-odd   
block components),  
the super-determinant is defined as 
\begin{equation}
\text{sdet} M = \frac{\text{det} (A-B D^{-1} C)}{\text{det} D}=\frac{\text{det}A}{\text{det}(D-CA^{-1}B)}, 
\end{equation}
and the supertrace is 
\begin{equation}
\text{str} M= \text{tr} A -\text{tr} D.
\end{equation}
(For more details, see Ref.\cite{dictionaryonsuperalgebras} for instance.)

\section{Lagrange Formalism}\label{sectlagrange}

As a supplement, we argue about the Lagrange formalism, which readily reproduces the results obtained in the Hamilton formalism. 
The one-particle Lagrangian is given by 
\begin{equation}
L=\frac{M}{2}(\eta_{ab}\dot{x}^a\dot{x}^b-\epsilon_{\alpha\beta}\dot{\theta}^{\alpha}\dot{\theta}^{\beta})+\dot{x}^a A_a+\dot{\theta}^{\alpha}A_{\alpha},
\end{equation}
with the constraint 
\begin{equation}
\eta_{ab}x^ax^b-\epsilon_{\alpha\beta}\theta^{\alpha}\theta^{\beta}=-1.
\end{equation}
In the LLL limit $M\rightarrow 0$, the Lagrangian is reduced to 
\begin{equation}
L_{eff}=\dot{x}^a A_a+\dot{\theta}^{\alpha}A_{\alpha}
=-iI\psi^{\dagger}k\frac{d}{dt}\psi,
\end{equation}
with $\psi$ (\ref{non-compactHopfspinor}) and $k$ (\ref{explicitk}). 
Regarding $\psi$ as the fundamental quantity, its canonical conjugate momentum is derived as 
\begin{equation}
\pi={\partial}L_{eff}/{ \partial \dot{\psi}}=-iI k\psi^*,
\end{equation}
where the right derivative was used. 
Imposing the commutation relations 
\begin{equation}
[\psi^A,\pi_B]_{\pm}=i\delta^A_{~B},
\end{equation}
the complex conjugation $\psi^*$ is represented as 
\begin{equation}
\psi^*=\frac{1}{I} k'\frac{\partial}{\partial\psi},
\label{effectivepsi*}
\end{equation}
with $k'$ (\ref{matrixk'}). 
Inserting (\ref{effectivepsi*}) to the non-compact SUSY Hopf map (\ref{complexSUSYhopfmap}), $x^a$ and $\theta^{\alpha}$ are represented as 
\begin{equation}
X^a=-\alpha\psi^t j^a\frac{\partial}{\partial\psi},
~~\Theta^{\alpha}=\alpha\psi^t j^{\alpha}\frac{\partial}{\partial\psi},
\end{equation}
which satisfy the super-hyperbolic fuzzy algebra (\ref{fuzzysuperhyperboloidalgebras}).
Similarly, the normalization condition (\ref{normalizationpsi}) is rewritten as 
\begin{equation}
\psi^t\frac{\partial}{\partial\psi} f_{LLL}=I f_{LLL},
\end{equation}
and it determines the LLL bases as in Eq.(\ref{explicitsusylllbases}).
 
\section{Irreducible Representation of $SU(1,1)$}\label{appendirredrepsu11}

Here, we summarize the irreducible representations of the $SU(1,1)$ group. (A more complete discussion is found  in Ref.\cite{am48(1947)}.) The irreducible representations are classified as  
(1) the principal discrete series, (2) the principal continuous series, and  
(3) the complementary continuous series. 
The principal discrete and continuous series form the complete bases.

The $SU(1,1)$ Casimir operator is given by the Hermitian operator 
\begin{equation}
\eta_{ab}L^a L^b=(L^x)^2+(L^y)^2-(L^z)^2,
\end{equation}
and its eigenvalues are real numbers that can be negative as well as positive. We express the eigenvalues as   
\begin{equation}
-l(l-1).
\label{casimirosp12}
\end{equation}
When $l$ is a real number, the eigenvalue satisfies  
\begin{equation}
-l(l-1)\le \frac{1}{4}.
\end{equation}
Meanwhile, when 
\begin{equation}
-l(l-1) > \frac{1}{4},
\end{equation}
$l$ can be parameterized as 
\begin{equation}
l=\frac{1}{2}+i\kappa
\label{principalcontinuous}
\end{equation}
with an arbitrary real constant $\kappa$, and (\ref{principalcontinuous}) provides $-j(j-1)=\frac{1}{4}+\kappa^2>\frac{1}{4}$.
The eigenvalue of  $L^z$ is given by a real number $m$, and simultaneous eigenstates of $\eta_{ab}L^aL^b$ and $L^z$ are introduced as 
\begin{subequations}
\begin{align}
&\eta_{ab}L^aL^b|l,m\rangle=-l(l-1)|l,m\rangle,\\
&L^z|l,m\rangle=m|l,m\rangle.
\end{align}
\end{subequations}
The raising and lowering operators are defined by
\begin{equation}
L^{\pm}=L^x\pm iL^y, 
\end{equation}
and yield relations 
\begin{subequations}
\begin{align}
&{L^+}^{\dagger} {L^+}=\eta_{ab}L^aL^b+(L^z)^2+L^z,\\
&{L^-}^{\dagger} {L^-}=\eta_{ab}L^aL^b+(L^z)^2-L^z. 
\end{align}\label{raisingandloweringrelations}
\end{subequations}
From the expectation values of (\ref{raisingandloweringrelations}) sandwiched by 
$|l,m\rangle$, the conditions for $l$ and $m$ are derived as 
\begin{subequations}
\begin{align}
&0 \le -l(l-1)+m(m+1),\\
&0 \le -l(l-1)+m(m-1).
\end{align}
\end{subequations}

\subsection{Principal Discrete Series}

With a real positive $l$
\begin{equation}
l >0, 
\end{equation}
two independent irreducible representations are introduced: 
\begin{align}
&m=l,l+1,l+2,\cdots,\label{discretepositive}\\ 
&m=-l,-l-1,-l-2,\cdots. \label{discretenegative}
\end{align}
(\ref{discretepositive}) and (\ref{discretenegative}) are named the positive and negative discrete series, respectively. 

\subsection{Principal Continuous Series}

When $l$ takes the form  (\ref{principalcontinuous}), the irreducible representation is specified as
\begin{equation}
|l,\alpha; m\rangle.
\end{equation}
Here, $m$ takes the form
\begin{equation}
m=\alpha,\alpha + 1,\alpha + 2, \cdots,
\end{equation}
or alternatively,
\begin{equation}
m=\alpha,\alpha- 1,\alpha - 2, \cdots,
\end{equation}
with $0 \le \alpha <1$.

\subsection{Complementary Continuous Series}

When $l$ satisfies the constraint 
\begin{equation}
l(l-1)<\alpha(\alpha-1)
\end{equation}
or
\begin{equation}
l-\frac{1}{2}< |\alpha-\frac{1}{2}|,
\end{equation}
with the parameters $0 \le \alpha <1$ and $1/2<l<1$, the irreducible representation is specified as  
\begin{equation}
|l,\alpha; m\rangle,
\end{equation}
where  $m$ takes the following values:    
\begin{equation}
m=\alpha,\alpha+ 1,\alpha+ 2, \cdots,
\end{equation}
or alternatively, 
\begin{equation}
m=\alpha,\alpha- 1,\alpha- 2, \cdots.
\end{equation}


\end{document}